# Expanding Active Matter to the Third Dimension: Exploring Short and Long-Range Particle-Wall Interactions.

Sandeep Ramteke[1], Jordan Dehmel[2], Touvia Miloh[3], Jarrod Schiffbauer[2,4], Alicia Boymelgreen[1]
[1]Dept of Mechanical and Materials Engineering, Florida International University, Miami, Florida
[2]Dept of Physics, Colorado Mesa University, Grand Junction, Colorado
[3]School of Mechanical Engineering, Tel Aviv University, Tel Aviv, Israel
[4]Dept of Mechanical Engineering, University of Colorado Boulder, Grand Junction, Colorado

Most active colloid experiments are quasi-2D. Here a 3D density-matched solution of active particles propelled and aligned with an AC electric field uniquely facilitates measurement of short and long-range particle-wall interactions. Near-wall mobility is reduced by Stokes drag and local electric-field distortion. Long-range attractions concentrate particles at upper and lower walls at a ratio dependent on particle orientation and rate proportional to speed and confinement. This approach may be extended to other active systems to understand particle-wall interactions and non-equilibrium phenomena.

The experimental realization of synthetic 3D active systems remains a significant challenge but is critical for modelling non-equilibrium statistical mechanics and complex systems [1,2], biological interactions [3,4], near [5,6], and long-range wall effects [7–9] and the optimization of diffusion driven applications (e.g., thermal transport [10]). Here we address this challenge by density matching active colloids and aligning the propulsive direction. The resultant experimental characterization of both short- and long-range particle-wall interactions provides unique insight into the limitations and assumptions of existing theoretical models and simulations, as well as a clear pathway to the application of this approach to other active systems.

In accordance with the scallop theorem [11], a key requirement for the translation of microscale active particles is symmetry-breaking. As in nature, this can be achieved through geometry (e.g., helical flagella [12]) but is more commonly attained by breaking material symmetry with a "Janus" [13] configuration; usually in the form of a polystyrene bead partially coated with a metal - the choice of which depends on the propulsive mechanism [14]. The high density of the metal causes particles to rotate the metallic hemisphere downwards (gyrotaxis) and sediment to the lower boundary [13–16]. If the colloidal activity is sufficient to counteract gyrotaxis and realign the metallo-dielectric interface with a component perpendicular to the wall, colloids exhibit stuck, oscillatory or sliding states at the lower boundary [15–19]. If the interface remains parallel to the wall, colloids may migrate towards to the upper boundary under gravitaxis [20], exhibiting the aforementioned states there. In both cases, the experimental system is quasi-2D.

Once at the wall, the particle mobility and equilibrium orientation are dictated by the complex interplay of short-range particle-wall interactions including increased Stokes drag [18,21,22], electrostatic forces [18], phoretic interaction [23], electrohydrodynamic flow [24,25], and amplified electrical or chemical [26,27] gradients. Wall accumulation is also compounded by long-range hydrodynamic attraction of swimmers to the walls [5,15,17,18].

Characterization of wall effects using external trapping (e.g., magnetic [28], optical [29,30]) suffers from the difficulty that the trapping mechanism may interfere with the interaction being tested. Mitigating gravitational effects by conducting experiments in microgravity [31,32] can be technically challenging [33], and cost-prohibitive, as the length of a typical experiment requires long-term microgravity such as on the International Space Station (ISS). A straightforward approach to counteract sedimentation is to increase the density of the suspending solution so that the JPs are neutrally buoyant [34]. This approach is not entirely comparable to microgravity since gyrotaxis can still modulate the mobility [15,35]. It has nonetheless recently been used to map the phase space of a concentrated suspension of metallo-dielectric JPs subject to AC electric fields [36] and measure the accumulation of catalytic colloids at the boundaries and their near wall-bound states [15].

In the present work, we observe both short- and long-range particle-wall interactions in a density-matched suspension of active Janus particles propelled by induced-charge electrophoresis (ICEP) in a uniform alternating current (AC) electric field. The particles are fabricated from polystyrene (Ps) spheres (radius $a = 2.5\mu m$, (Fluro-Max, Thermo Scientific) on which a thin 15nmTi/15nmAu hemispherical coating is deposited following the procedure in [37] (S.1 in [38]). The particles are density matched at ambient lab temperature of 25°C in a 0.70 w/w glycerol (Sigma Aldrich, $\rho_g = 1.26\,\text{g/cm}^3$ [39], 92.09 g/mol, $C_3H_8O_3$) to deionized water ($\rho_w =$

0.997 g/cm³) mixture with density $\bar{\rho}_f = 1.17 \text{g/cm}^3$ [39]. The relatively low glycerol fraction required is attributed to the nonuniformity of the coating at the equator, which may be modeled as elliptical rather than hemispherical [16] (Figure 1b-c, S.2 in [38]). The density-matched solution is placed inside single ($2H = 120\mu m$) or stacked ($2H = 240\mu m$) silicon reservoirs (Secure Seal Space) that are sandwiched between an Indium Tin Oxide (ITO) coated glass slide (Delta Technologies) and ITO-coated cover slip (SPI). To minimize particle sticking to the ITO [35,40,41], a 20nm $SiO_2$ layer is thermally deposited by E-beam on the ITO-coated side of both the glass slide and cover-slip. In Figure 1d-e, a z-stack imaged with a spinning disc confocal microscope (Nikon CSU-X1 mounted on Nikon TI2-E) and motorized stage (Ti2-S-SE-E) with a 20x objective illustrates the difference in distribution of particles in DI water, which is sedimentation dominated and the 0.70w/w glycerol/DI mixture. The ITO-coated substrates serve as both channel walls and electrodes which are connected to an AC function generator (Keysight, 33210A) for voltages between 3.3-16 $V_{PP}$ and frequencies ranging from 0.2kHz to 5kHz. Image capture is facilitated by a high-resolution (ORCA-Fusion C14440-20UP) at a rate of 4fps. Mobility analysis was conducted using Image J with the speckle tracker plugin [42] and post processed with in-house Python code [43].

The advantage of using an AC electric field to drive the active colloids is two-fold. Firstly, the particle velocity scales with the radius [44] such that larger, more buoyant particles (Figure 1c) can be used (in contrast to catalytically driven colloids whose speeds are inverse to the particle radius [45]). More critically, gravitaxis, preventing bulk mobility measurement [15], is inherently mitigated by the field-induced electrostatic and electrohydrodynamic torques which can largely counteract gravitational torque to align the metallo-dielectric interface with the electric field [35]. Thus, the velocity $U$ in the x-y plane dominates that in the z direction facilitating measurement of $U(z)$ (Figure 1f, Figures 2-3) and accumulation at the boundaries as a function of time and confinement (Figure 4). Given the strongly temperature dependent properties of glycerol [39,46], a key concern with the application of the electric field is Joule heating causing viscosity gradients and electrothermal flow [47]. Here, we aim to minimize these effects by using DI water with the glycerol which yields a solution conductivity less than $1\mu S/cm$ [48]. The temperature of the device under a continuous applied field is measured externally with an infrared camera and internally with temperature sensitive fluorescent dyes ([48,49])over a period of 10 minutes with little variation observed (S3 in [38]); similarly there is no marked increased in JP velocity over time which would correlate with reduced viscosity (S3 in [38]). No background flow (buoyancy driven convection) is observed corresponding with an estimated Rayleigh number well below the critical value for emergence of Rayleigh-Bénard instability. Combined, these observations suggest that any Joule heating effects are secondary to the dominant mechanisms described below.

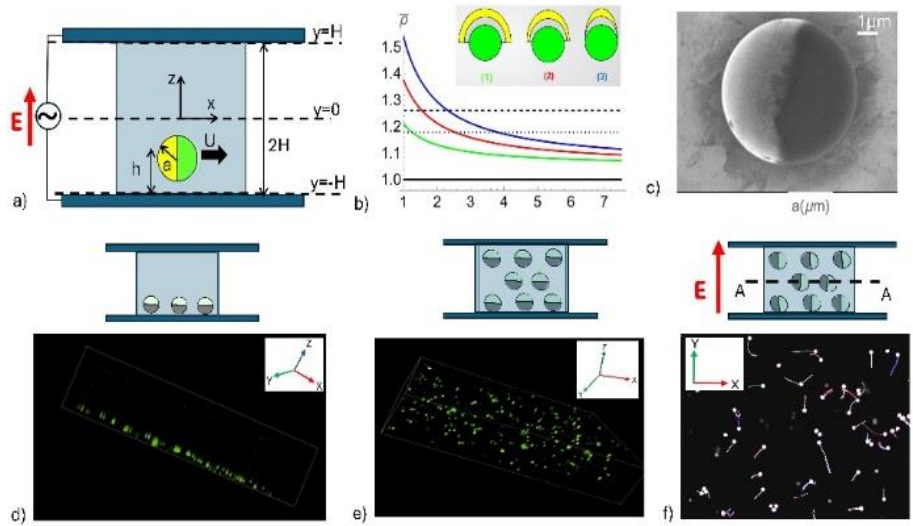

Figure 1: a) Schematic of experimental setup and coordinate system, b) Plot of JP density as a function of size for hemispherical (1) and ellipsoidal (2)-(3) coatings. Dashed, dotted and solid lines denote densities estimated at 25°C of glycerol, 0.70w/w glycerol/water mixture and pure water [38]. c) SEM image of Janus particle exhibiting non-uniformities around the equator resulting in reduced density. d) JPs in pure DI water sediment e) Density-matched JPs are suspended throughout the bulk. f) Cross sectional image taken in the center of the chamber with particle tracks superimposed showing translation in the x-y plane.

The physical mechanism driving the Janus particles is induced-charge electrophoresis (ICEP) [37,44,50] which relies on the formation of an induced electric double layer (EDL) at the polarizable

*Contact author: aboymelg@fiu.edu

(metallic) hemisphere, whose motion under the same electric field yields induced-charge electroosmotic flow (ICEO). The broken symmetry of the Janus particle, where flow is only generated at the metallic hemisphere, results in velocity $U$ in the $x_1$ direction with the dielectric hemisphere forward (Figure 1a). Following [50] (see also Appendix), in the absence of wall effects, the dimensionless theoretical ICEP force (normalized by $\varepsilon\varepsilon_0 a^2 E_0^2$ where $\varepsilon$ is the relative permittivity of the fluid, $\varepsilon_0$ the permittivity of a vacuum and $E_0$ the amplitude of the applied electric field) for a single particle translating in the bulk is given by:

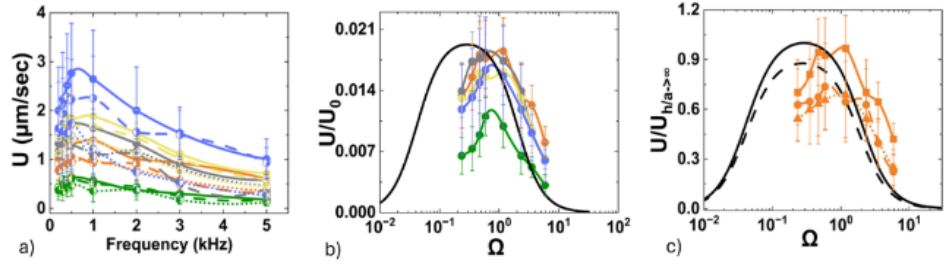

Figure 2: a) Experimental measurement of velocity in the bulk (solid lines), lower (dashed) and upper (dotted) boundaries in a channel $2H = 120\mu m$. Colors correspond to values of $E_0$(V/m): Blue ($2.5E^9$), Yellow ($1.74E^9$), Grey ($1.41E^9$), Orange($1.1E^9$), Green ($8.5E^8$)b) Non-dimensional theoretical and experimental velocities at $z=0$; fitting factor of $\Lambda = 1/7$ is applied to the theoretical model (solid black line, Eq. 3 with a/h=0.04). c) Comparison of near-wall and bulk mobility normalized by the maximum velocity in the bulk. Dashed black line corresponds to Eq. 3 with a/h=0.95.

$$F_{ICEP}^{(1)} = \frac{27\pi}{16}\left[\frac{\Omega_H^2}{1+\Omega_H^2}\right]\left[\frac{1}{4+\Omega^2}\right] \quad (1)$$

where the factor of $1/(4+\Omega^2)$ (where $\Omega = \omega a \lambda_0/D$; $\omega$ is the frequency of the applied field, $\lambda_0$, the Debye length and $D$ the diffusivity of a z:z symmetric solute) accounts for the high frequency screening due to relaxation of the EDL at the metallic hemisphere while $\Omega_H^2/(1+\Omega_H^2)$ (where $\Omega_H = \omega H \lambda_0/D$; $2H$ is the chamber height) accounts for the low frequency decay due to screening of the electrodes [37].

The modification of the ICEP force due to wall proximity can be obtained following the asymptotic approach of [37,51] where we have extended our solution [37] to the fifth order in $a/h$. Based on the observation that the ratio of the velocity near the wall relative to the bulk is a function of frequency (Figure 2a); a higher order correction accounting for the conducting nature of the wall/electrode is included following [24] (see Appendix) such that

$$F_{ICEP}^{(1)} = \frac{27\pi}{16}\left[\frac{\Omega_H^2}{1+\Omega_H^2}\right] \times$$
$$\left[\frac{\left(1-\frac{9}{16}\left(\frac{a}{h}\right)+\frac{1}{8}\left(\frac{a}{h}\right)^3-\frac{45}{256}\left(\frac{a}{h}\right)^4-\frac{1}{16}\left(\frac{a}{h}\right)^5\right)^{-1}}{4-\frac{1}{2}\left(\frac{a}{h}\right)^3+\frac{1}{16}\left(\frac{a}{h}\right)^5+\Omega^2\left(1-\frac{1}{4}\left(\frac{a}{h}\right)^3-\frac{1}{8}\left(\frac{a}{h}\right)^5\right)}\right] \quad (2)$$

which reduces to Eq. (1) in the remote limit, namely when $a/h \to 0$. The resultant expression indicates that the ICEP force increases with proximity to the channel wall and as a function of the frequency of the applied field. The corresponding velocity $U$ for a freely suspended particle translating in the $x_1$ direction is derived by equating $F_{ICEP}^{(1)}$ (Eq. (2)) with the Stokes drag defined according to [52] such that

$$U = \frac{F_{ICEP}^{(1)}}{6\pi f};$$
$$f = 1-\frac{9}{16}\left(\frac{a}{h}\right)+\frac{1}{8}\left(\frac{a}{h}\right)^3-\frac{45}{256}\left(\frac{a}{h}\right)^4-\frac{1}{16}\left(\frac{a}{h}\right)^5 \quad (3)$$

Note that the natural scaling for the velocity is $U_0 = \varepsilon\varepsilon_0 E_0^2 a/\eta$ where $\eta$ denotes the dynamic viscosity.

In accordance with Eq. (3), the velocity measured near the wall is consistently smaller than that in the bulk (Figure 2a). As the magnitude of the applied field increases, an asymmetry arises between the velocities in the upper and lower bounds. This may be attributed to increased particle sticking that occurs at high voltages and long durations. To validate that the propulsive mechanism is ICEP rather than electrothermal flow [48], the experimental velocity measurements are normalized by the characteristic ICEP velocity, $U_0 = \varepsilon\varepsilon_0 E^2 a/\eta$ and plotted against the non-dimensional characteristic frequency $\Omega$ in Figure 2b. The collapse of the curves (except for the lowest voltage) indicates that the mobility scales quadratically with the applied field rather than to the fourth power which would be expected for electrothermal flow [48,53]. The departure of this scaling for the smallest applied field could be due to gyrotaxis [35]. To obtain the characteristic length scales, the viscosity of a 0.70w/w glycerol/water solution at ambient lab temperature (25°C) was measured in a viscometer (Brookfield AMETEK DV2T) as $\eta = 0.017 kg/ms$ in agreement with published standards [46]. The diffusivity is inferred from the viscosity based on Stokes-Einstein formula as $D = 1.15 \cdot 10^{-10}$m$^2$/s and the solute permittivity $\varepsilon$ is estimated from [54,55] and given as $\varepsilon = 53.5$ and $\varepsilon_0 = 8.67 \cdot 10^{-12}$F/m (see also [38]). The thickness of the Debye layer $\lambda_0 \sim 10nm$ is used as a fitting factor. Although smaller than expected, such discrepancy is widely observed in ICEP and ICEO experiments and has previously been attributed to the capacitance of the Stern layer and high voltage effects [56,57]. Similarly, the theoretical model over predicts the velocity by approximately an order of magnitude. Accordingly, a fitting factor of $\Lambda = 1/7$ is used to modify the magnitude of the ICEP force [52,53]. Importantly, the 3D nature of the system uniquely allows us to validate that these quantitative discrepancies in the frequency dispersion exist both in the bulk and near the wall demonstrating that this is not solely a wall effect but an electrokinetic one. In Figure 2c, we compare the near wall and bulk mobilities at a constant electric field ($E_0 = 1.1 \cdot 10^9 V/m$) to the theoretical model described in Eq. (3), evaluated at an estimated height of $h/a = 1.05$. The overprediction of the near wall velocity implies that $h/a < 1.05$ and highlights the need for a higher order model or numerical simulations valid in the near- contact limit for quantitative agreement.

*Contact author: aboymelg@fiu.edu

In Figure 3, we compare the velocity of the Janus particles for varying values of $h$ in chambers of $2H = 120 \mu m$ (Figure 3a) and $2H = 240 \mu m$ (Figure 3b) at a constant electric field $E_0 = 1.1 \cdot 10^9 V/m$. The velocities at each plane are normalized by the mobility at the channel centerline $(z = 0)$. The maxima in Figure 3a - compared to the expected plateau observed in Figure 3b - raises the question as to whether confinement restricts the bulk mobility even for electrode spacing an order of magnitude larger than the radius ($H = 24a$). Replotting the normalized velocities as a function of the height scaled with the radius, $h/a$, it is observed that a constant velocity is approached at $h/a > 20$ suggesting the peak in Figure 3a does reflect the maximum bulk mobility (Figure 3c). The frequency dependence of the ratio $U_{z=0}/U_{z\max}$

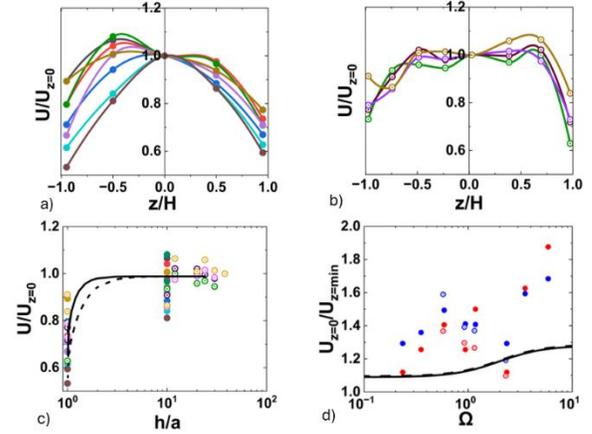

Figure 3: Variation of mobility normalized by the midplane velocity ($U_{z=0}$) as function of chamber height; a) $2H=120$μm and b) $2H=240$μm. c-d) Variation of the ratio between near wall to bulk mobility as a function of c) $h/a$ and d) $\Omega$. Solid circles correspond to 2H=120um, open circles to 2H=240um. Colors denote frequencies according to black (0.2kHz), red (0.3kHz), blue (0.4kHz), green (0.5kHz), purple(1kHz), mustard(2kHz), brown (5kHz). Lines correspond to Eq.1.3 for $\Omega$=0.1 (solid) and $\Omega$=10 (dashed)

(where $z_{\max} = \pm(H-h)$) is compared to the theoretical model Eq. (3) in Figure 3d. The model qualitatively captures the initial increase of $U_{z=0}/U_{z\max}$ with frequency but does not predict the subsequent decrease, which may be attributed either to overlapping EDL effects at the wall or the frequency shift between near wall and bulk mobilities previously predicted by numerical simulations [37].

Overall, the persistence of wall effects deep into the chamber is surprising. Even for relatively large frequencies $(\Omega = 10)$, where the decay of wall effects occurs more slowly (Figure 3c), the velocity should approach 98% of the bulk for $h/a > 5$. That the velocity is suppressed well beyond this limit at both boundaries implies the presence of long-range particle-wall interactions that must be accounted for. The unique 3D nature of our density matched solution and control of mobility through the electric field strength facilitates measurement of the time-dependent accumulation at the boundaries as a function particles mobility and orientation (Figure 4) - previously only examined theoretically and numerically (e.g., [5,15,17,18]). The probability that a particle will be in a position $p(z)$ at a given time $t$, is calculated according to $p(z) = N_z/\sum_{i=1}^{i=N} N_i$ where $N_z$ is the number of particles at a specific height and $N$ the total number of layers. Here, we measure $N_z$ at 25μm intervals such that $N = 2H/25$. It is noted that since the field of view is restricted, the value of the denominator can vary over time as particles come in and out of frame. However, we find that this number mostly remains constant with maximum variation as the voltage is increased and particles travel more rapidly in and out of the $x-y$ plane (see S.4 in [38]). It is readily observed that as the applied field increases, the rate of accumulation at the top and bottom substrates (and conversely, the depletion in the bulk) occurs faster and saturation is achieved at shorter times (Figure 4a). The asymmetry between the upper and lower boundaries (Figures 4a, b), is explained by the gravitational torque which rotates the metallic hemisphere downward, resulting in a velocity component in the positive z-direction [35] (see S.5 in [38]). Accordingly, the asymmetry is more pronounced at lower voltages where the electrostatic and electrohydrodynamic torques are weaker [35]. Interestingly, increasing the height of the chamber does not change the steady state distribution but only the rate at which saturation occurs (Figure 4c) – decreasing with increasing height. This again corresponds qualitatively with [8] where in the limit of relatively large spacing, the variation in saturation distribution is small – although we also note that the experimental Peclet numbers are much larger than the ones for which the large channel limit is valid such that the problem, in fact remains an open question.

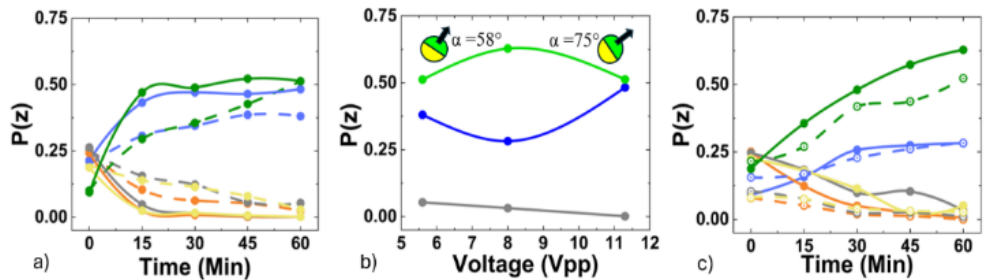

Figure 4: a) *p(z)* at varying heights (Blue ($h/H\sim -1$), Orange($h/H\sim -0.5$),Grey($h/H\sim 0$), Yellow($h/H\sim 0.5$), Green($h/H\sim 1$)for varying applied fields (dashed (5.66V), solid ( 11.3V)). b) *p(z)* as a function of applied field, height indicated by color. c) *p(z)* in chambers of height 2H=120μm (solid lines) and 2H=240μm (dashed lines).

*Contact author: aboymelg@fiu.edu

These results highlight the importance of experimental validation of such simulations which may only be achieved in a 3D system. They also feature the role of gravitational torque in active colloid mobility, which cannot be mitigated through density matching but requires an additional forcing to control the alignment of the active colloid. Recognizing this, we note that extension of the present density-matching approach to other active colloidal systems which do not constrain the orientation of the metallodielectric interface- such as the common catalytic Janus particles – requires the addition of an external aligning force (e.g., magnetic). Such experiments could provide deeper insight into short-range particle-wall interactions and the role of local chemical field gradients in modulating mobility. One can also envision a system where these properties are exploited to actively regulate localized particle volume fractions and orchestrate collective motion within active colloidal assemblies. Understanding and harnessing these capabilities are essential for enhancing the stability and performance of active colloidal dispersions or emulsions, thereby advancing practical applications in materials science and soft matter technologies.

**ACKNOWLEDGMENTS**
The authors are grateful for constructive discussions with Dr William Meyer regarding the material properties of glycerol-water solutions. The authors acknowledge the support of Dr Jay Bieber and Dr Robert Tufts at USF for the fabrication of the Janus particles. AB acknowledges the support of NSF-CASIS award #2126479. JS acknowledges the support of NSF-CASIS awards #2126451 and #2430509 and TM that of ISF 3858/24. AB and JS also acknowledge support by the International Space Station National Laboratory, managed by the Center for the Advancement of Science in Space, under user agreement number UA-2021-8340

**APPENDIX**

The theoretical model is based upon the weak- field formulation [58] wherein the governing electrostatic and hydrodynamic equations are linearized and decoupled. This technique is applicable when the electric potential $\phi$ is considered small with respect to thermal scale, $\varphi_t = k_B T/ze$, where $k_B$ is the Boltzmann constant, $T$ the temperature, $z$ the valence and $e$ the elementary charge. Previously, this model has been used to effectively model the ICEP velocity of metallo-dielectric Janus particles. Here, following Ristenpart et al. [24], we extend our previous work in [37] to account for the constant potential prevailing at the electrode/channel wall, which yields a higher- order, frequency dependent correction to the ICEP mobility in accordance with experimental observations.

Consider two electrodes (wall) where the potential on the upper wall $(z = H)$ is $\phi(H) = 0$ and that on the lower wall $(z = -H)$ is $\phi(H) = \Delta\phi_0$
Following [24] (Eq. 26), the ambient electric potential in the absence of the particle, can be expressed as

*Contact author: aboymelg@fiu.edu

$$\phi(z) = \frac{\Delta\phi_0}{2} e^{-i\omega t} \times$$
$$\left[1 - \frac{\sinh(\gamma z/H)\csc(\gamma) - i(z/H)\gamma\upsilon\coth(\gamma)}{1 - i\gamma\upsilon^2\coth(\gamma)}\right] \quad (A.1)$$

where $\upsilon^2 = \omega\lambda_0^2/D$ and $\gamma^2 = (H/\lambda_0)^2 - i(\omega H^2/D)$.
For a thin electric double layer (EDL), $\gamma \sim H/\lambda_0 \gg 1$ and thus near the upper wall, $z \sim H$ such that

$$\phi(z) = \frac{\Delta\phi_0}{2}\left[1 - \frac{1 - i(z/H)\gamma\upsilon^2}{1 - i\gamma\upsilon^2}\right] \quad (A.2)$$

where $\Omega_H = \gamma\upsilon^2 \sim \omega\lambda_0 H/D$

Next, consider a Janus particle lying at a distance $h$ from the upper wall. A Cartesian coordinate system attached to the JP $(x_1, y_1, z_1)$, admits $x_1 = x, y_1 = y, z_1 = -H + h + z$

Thus, in the absence of the JP, (A.2) can be written as

$$\phi(z_1) \simeq \frac{\Delta\phi_0}{2}\frac{\Omega_H}{\Omega_H + i}\left[\frac{h}{H} - \frac{z_1}{H}\right] \quad (A.3)$$

Using the dipole approximation for a dipole (including its first wall image)

$$\phi(z_1) \simeq \frac{\Delta\phi_0}{2}\frac{\Omega_H}{\Omega_H + i} \times$$
$$\left[\frac{h}{H} - \frac{1}{H}\left(z_1 + \frac{a^3 D z_1}{R_1^3} + \frac{a^3 D z_2}{R_2^3}\right)\right] \quad (A.4)$$

such that D denotes the corresponding dipole strength and $x_2 = x_1, y_2 = y_1, z_2 = z_1 - 2h$, $R_1^2 = x_1^2 + y_1^2 + z_1^2$, $R_2^2 = x_1^2 + y_1^2 + (z_1 - 2h)^2 = R_1^2 - 4z_1 h + 4h^2$. Note that $\phi(z_1 = h) = \phi(z = H) = 0$. Next, using the following asymptotic expansion in $a/h < 1$: $1/R_2^3 \sim 1/(2h)^3[1 - \frac{3}{2}(a/2h)^2 + ...]$, one gets

$$\phi(z_1) \simeq \frac{\Delta\phi_0}{2}\frac{\Omega_H}{\Omega_H + i} \times$$
$$\left[\begin{array}{l}\frac{h}{H} + \frac{a^3 D}{H(2h)^2}[1 - \frac{3}{2}(a/2h)^3 + ...] \\ -\frac{1}{H}\left(z_1 + \frac{a^3 D z_1}{R_1^3} + \frac{a^3 D z_1}{(2h)^3}(1 - \frac{3}{2}(a/2h)^3 + ...)\right)\end{array}\right]. \quad (A.5)$$

At the surface of the Janus particle, we impose the common RC boundary condition [37] for AC fields:

$$a\frac{\partial\phi}{\partial R_1}\bigg|_{R_1 = a} = -i\Omega(\phi - \phi_0); \quad \Omega = \frac{\omega\lambda_0 a}{D} \quad (A.6)$$

where when expressed in a JP-attached spherical coordinates $(R_1, \theta, \varphi)$, renders

$$\phi - \phi_0 \simeq -\frac{\Delta\phi_0}{2}\frac{\Omega_H}{\Omega_H + i} \times$$
$$\left[R_1 + \frac{a^3 D}{R_1^2} + \frac{a^3 D R_1}{(2h)^3}(1 - \frac{3}{2}(a/2h)^2 + ...)\right]\sin\theta\cos\varphi$$
(A.7)

which renders

$$D \simeq \left(\frac{2 - i\Omega}{1 + i\Omega} - \left(\frac{a}{2h}\right)^3[1 - \frac{3}{2}(\frac{a}{2h})^2]\right)^{-1} \quad (A.8)$$

Note that for $a/h \to \infty$ and $\Omega = 0$, one gets $D = 1/2$ in accordance with [37].

To evaluate the ICEP force exerted on a polarizable colloid ($Q$ and $\chi$ represent the induced-charge and ambient potential), we use the Teubner formulation [58] which for a thin EDL renders [37].

$$F_1^{ICEP} = \frac{3T}{16}\left(\frac{\lambda}{\lambda_0^2}\right)\int_S (\delta_{ij} - n_i n_j)Q^* \frac{\partial\chi}{\partial x_j}dS \quad (A.9)$$

Here S denotes the surface of the metallic hemisphere, **n** is its normal vector and (*) represents the complex conjugate. The geometrical parameter T in (A.9), accounts for the proximity effect of the nearby no-slip wall which following Davis and Crowdy [22], can be estimated to be fifth-order in wall spacing as

$$T \sim \left[1 - \frac{9}{16}\left(\frac{a}{h}\right) + \frac{1}{4}\left(\frac{a}{h}\right)^3 - \frac{63}{256}\left(\frac{a}{h}\right)^4 - \frac{1}{16}\left(\frac{a}{h}\right)^5\right]^{-1} \quad (A.10)$$

Taking advantage of the axial symmetry around $x_1$ of the potential $\chi(r, \theta)$, one gets

$$\frac{\partial\chi}{\partial x_1} - n_1\frac{\partial\chi}{\partial r} = \frac{\partial\chi}{\partial\theta}\frac{\partial\theta}{\partial x_1} \quad (A.11)$$

and

$$\frac{\partial\theta}{\partial x_1} = -\frac{\sin\theta}{a} \quad (A.12)$$

Finally, noting that at the metallic hemisphere $S$, $(\phi = 0)$ and thus [58] $0 = -(\lambda/\lambda_0)^2 Q + \chi$, we obtain

$$F_1^{ICEP} = \frac{3}{16a}T\int_S \chi^* \frac{\partial\chi}{\partial r}\sin\theta dS \quad (A.13)$$

Recalling next that $\chi(r, \theta) = \phi(z_1) - \phi_0$, one readily finds from (A.7) that

$$\frac{1}{2}\chi(a, \theta) \simeq -\frac{\Delta\phi_0}{2H}\frac{\Omega_H}{\Omega_H + i}a \times$$
$$\left[1 + D\left(1 + \left(\frac{a}{2h}\right)^3 + ...\right)\sin\theta\cos\varphi\right] \quad (A.14)$$

*Contact author: aboymelg@fiu.edu

and

$$\frac{1}{2}\frac{\partial \chi}{\partial \theta}(a,\theta) \simeq -\frac{\Delta\phi_0}{2H}\frac{\Omega_H}{\Omega_H+i}a \times \left[1+D\left(1+\left(\frac{a}{2h}\right)^3+...\right)\cos\theta\cos\varphi\right] \quad (A.15)$$

Substituting the above in Eq. (A.13) yields

$$F_1^{ICEP} \simeq \left(\frac{\Delta\phi_0}{2H}\right)^2 \frac{\Omega_H^2}{\Omega_H^2+1}T\left|1+D\left(1+\left(\frac{a}{2h}\right)^3+...\right)\right|^2 \times$$

$$\frac{3}{4}a^3\int_0^{2\pi}\cos^2\varphi d\varphi\int_0^{\pi/2}\sin^3\theta\cos\theta d\theta. \quad (A.16)$$

or

$$F_1^{ICEP} \simeq \frac{3\pi}{16}a^3\left(\frac{\Delta\phi_0}{2H}\right)^2\frac{\Omega_H^2}{\Omega_H^2+1} \times$$

$$T\left|1+D\left(1+\frac{1}{8}\left(\frac{a}{h}\right)^3+...\right)\right|^2 \quad (A.17)$$

Finally, Substitution of Eqs. (A.7) and (A.9) into (A.16) and gathering terms to $(a/h)^5$, we finally obtain

$$F_1^{ICEP} = \frac{27\pi}{16}a^3\left(\frac{\Delta\phi_0}{2H}\right)^2\frac{\Omega_H^2}{\Omega_H^2+1} \times$$

$$\frac{T}{4-\frac{1}{2}\left(\frac{a}{h}\right)^3+\frac{1}{16}\left(\frac{a}{h}\right)^5+\Omega^2\left(1+\frac{1}{4}\left(\frac{a}{h}\right)^3-\frac{1}{8}\left(\frac{a}{h}\right)^5\right)} \quad (A.18)$$

Note that for the unbounded case, $(a/h) \to 0$ and $T \to 1$, $F_1^{ICEP}$ reduces to

$$F_1^{ICEP} = \frac{27\pi}{16}a^3\left(\frac{\Delta\phi_0}{2H}\right)^2\frac{\Omega_H^2}{\Omega_H^2+1}\cdot\frac{1}{4+\Omega^2}\cdot \quad (A.18)$$

which is identical to Eq.15 in Ref. [37] where $\Delta\phi_0/2H \sim E_0$ and $a=1$.

*Contact author: aboymelg@fiu.edu

*Contact author: aboymelg@fiu.edu